\documentclass[aps,pra,twocolumn,floatfix]{revtex4}
\usepackage{bm}
\usepackage{epsfig}

\begin{document}

\title{Intrinsically faint quasars: evidence for $meV$ axion
dark matter in the Universe}
\author{Anatoly A. Svidzinsky}
\affiliation{Department of Physics, Institute for Quantum Studies,
Texas A\&M University, TX 77843-4242}

\date{\today }

\begin{abstract}
Growing amount of observations indicate
presence of intrinsically faint
quasar subgroup (a few \% of known quasars)
with noncosmological quantized redshift.
Here we find an analytical
solution of Einstein equations describing bubbles made from axions with
periodic interaction potential. Such particles are currently considered as
one of the leading dark matter candidate. The bubble interior possesses
equal gravitational redshift which can have any value between zero and
infinity. Quantum pressure supports the bubble against collapse and yields
states stable on the scale more then
hundreds million years. Our results explain
the observed quantization of quasar redshift and suggest that intrinsically
faint point-like quasars associated with nearby galaxies are axionic bubbles
with masses 10$^8$-10$^9M_{\odot }$ and radii 10$^3$-10$^4R_{\odot }$. They
are born in active galaxies and ejected into surrounding space. Properties
of such quasars unambiguously indicate presence of axion dark matter in the
Universe and yield the axion mass $m\approx 1$ meV, which fits in the open
axion mass window constrained by astrophysical and cosmological arguments.
\end{abstract}

\maketitle

Based on observations, Karlsson \cite{Karl90} has noted division of quasars
(QSOs) into two groups with different redshift properties and concluded the
following. If we select QSOs associated with
most nearby (distance $d\lesssim 50-100$ Mpc), galaxies then their
redshift is close to certain values (quantized), as shown in Fig. 3 below.
Meanwhile, in QSO samples associated with distant
galaxies no periodicity in intrinsic redshift is observed.
Such a division is supported by later studies
of QSOs associated with most nearby galaxies where the quantization was
confirmed \cite{Arp90,Burb01} and distant ($0.01<z_{\text{gal}}<0.3$) galaxies
for which absence of any periodicity was claimed \cite{Hawk02}.
The observations
suggest existence of intrinsically faint (optical luminosity
$L=10^5-10^7L_{\odot }$) QSO subgroup with quantized noncosmological
redshift. Being intrinsically faint, such objects are not detected from large
distances (which yields disappearance of redshift quantization in distant QSO
samples) and constitute only a few \% of the known QSO population.
Observations indicate that such quasars are ejected from nearby active
galaxies or in the process of ejection from the galactic
nucleus \cite{1,Arp87}.

Here we show that bubbles of dark matter with periodic interaction
potential, masses about $10^8-10^9M_{\odot }$ and radii $10^3-10^4R_{\odot }$
can explain the intrinsically faint quasars. The bubble is supported against
collapse by quantum pressure and decays on a time scale more than hundreds
million years.  Hypothetical axions, one of the leading dark matter candidate,
fit well into this picture and can account for the redshift quantization. Usual
baryonic matter falls into the bubble interior, heated by the release of the
gravitational energy and produce electromagnetic radiation that freely
propagate into surrounding space.

In this Letter we study massive real scalar field $\varphi $ with periodic
interaction potential
\begin{equation}
\label{p1}V(\varphi )=V_0[1-\cos (\varphi /f)],
\end{equation}
where $V_0>0$. This potential is quite general and derived in quantum filed
theory in connection with pseudo Nambu-Goldstone bosons (PNGBs) \cite{Hill88}%
. In all such models, the key ingredients are the scales of global symmetry
breaking $f$ and explicit symmetry breaking $(V_0)^{1/4}$. One of the
examples of a light hypothetical PNGB is the axion which possess
extraordinarily feeble couplings to matter and radiation and is
well-motivated dark matter candidate \cite{Brad03}. If the axion exists,
astrophysical and cosmological arguments constrain its mass to be in the
range of $m=10^{-6}-3\times 10^{-3}$ eV and the global symmetry-breaking scale
to lie in a window $f\approx 10^7$ GeV $\times 0.62/m\text{(eV)}=2\times
10^{9}-6\times 10^{12}$ GeV \cite{Brad03}.

We consider spherically symmetric system with metric
\begin{equation}
\label{b1}ds^2=-N^2dt^2+g^2dr^2+r^2d\Omega ^2,
\end{equation}
where $g$, the radial metric, and $N$, the lapse, are functions of $t$ and $%
r $ with $r$ being the circumferential radius. We introduce dimensionless
coordinates and define the unit of distance, time and $\varphi $ as
\begin{equation}
\label{a2}r_0=\frac \hbar {mc},\quad t_0=\frac \hbar {mc^2},\quad \varphi
_0=\frac 1{\sqrt{4\pi G}},
\end{equation}
where $c$ is the speed of light, $G$ is the gravitational constant, $m=\sqrt{%
V_0}/f$ is the particle mass. In dimensionless units the static Klein-Gordon
and Einstein equations describing the self-gravitating field $\varphi $ and
the metric are \cite{Seid90}
\begin{equation}
\label{b5}\frac{\varphi ^{\prime }}{g^2}\left( \frac{g^2+1}r-2rg^2V\right) +%
\frac{\varphi ^{\prime \prime }}{g^2}-\frac{\partial V}{\partial \varphi }%
=0,
\end{equation}
\begin{equation}
\label{b6}N^{\prime }=\frac N2\left[ \frac{g^2-1}r+r\left( \varphi ^{\prime
2}-2g^2V\right) \right] ,
\end{equation}
\begin{equation}
\label{b7}g^{\prime }=\frac g2\left[ \frac{1-g^2}r+r\left( \varphi ^{\prime
2}+2g^2V\right) \right] ,
\end{equation}
with boundary conditions
$$
g(0)=g(\infty )=N(\infty )=1,\quad
$$
$$
g^{\prime }(0)=N^{\prime }(0)=\varphi ^{\prime }(0)=0,\quad V(\varphi
(\infty ))=0,
$$
where prime denotes $\partial /\partial r$,
\begin{equation}
\label{a3}V=\frac 1{\alpha ^2}[1-\cos (\alpha \varphi )],\quad \alpha =\frac
1{\sqrt{4\pi G}f}=\frac{m_{\text{pl}}}{\sqrt{4\pi }f}
\end{equation}
is the dimensionless potential and the coupling parameter respectively, $m_{%
\text{pl}}=\sqrt{\hbar c/G}=1.2\times 10^{19}$ GeV is the Planck mass. The
interaction potential has degenerate minima at $\varphi =2\pi n/\alpha $,
where $n$ is an integer number. Here we show that in the limit of strong
nonlinearity, $\alpha \gg 1$, Eqs. (\ref{b5})-(\ref{b7}) have an approximate
static solution that describes a spherical bubble with surface width much
smaller then its radius $R$. The bubble surface is an interface between two
degenerate vacuum states with $\varphi =2\pi n/\alpha $ ($r<R$) and $\varphi
=0$ ($r>R$).

Outside the bubble Eqs. (\ref{b5})-(\ref{b7}) lead to the known Schwazschild
solution
\begin{equation}
\label{s1}g^2=\frac 1{1-2M/r},\quad N^2=1-\frac{2M}r,
\end{equation}
where $M$ is the bubble mass in units of $m_{\text{pl}}^2/m$.

Let us assume that $R\gg \alpha \gg 1$. Then, near the surface one can omit
terms with $1/r$ in Eqs. (\ref{b5})-(\ref{b7}) and take $r\approx R$, we
obtain
\begin{equation}
\label{b8}-2R\varphi ^{\prime }V+\frac{\varphi ^{\prime \prime }}{g^2}-\frac{%
\partial V}{\partial \varphi }=0,
\end{equation}
\begin{equation}
\label{b9}N^{\prime }=\frac{NR}2\left( \varphi ^{\prime 2}-2g^2V\right) ,
\end{equation}
\begin{equation}
\label{b10}g^{\prime }=\frac{gR}2\left( \varphi ^{\prime 2}+2g^2V\right) .
\end{equation}
Eqs. (\ref{b8})-(\ref{b10}) can be solved analytically. Their first integral
is
\begin{equation}
\label{b11}N=const,\quad \varphi ^{\prime 2}=2g^2V,\quad g^{\prime
}=Rg\varphi ^{\prime 2}.
\end{equation}
We assume $\varphi (0)=2\pi n/\alpha $, where $n=1,2,3,\ldots $ is the
number of kinks at the bubble surface, and $\varphi (r)$ monotonically
decreases with $r$. Eqs. (\ref{b11}) yield
\begin{equation}
\label{e4}\frac 1g=1-R\int_\varphi ^{\varphi (0)}\sqrt{2V}d\varphi ,\quad
\varphi ^{\prime }=-\frac{\sqrt{2V}}{1-R\int_\varphi ^{\varphi (0)}\sqrt{2V}%
d\varphi }.
\end{equation}
For $V(\varphi )$ given by Eq. (\ref{a3}) the final solution is%
$$
\frac{4R}{\alpha ^2}\ln |\sin (\alpha \varphi /2)|+\left[ 1-\frac{4R}{\alpha
^2}(2m-1)\right] \text{arctanh}[\cos (\alpha \varphi /2)]
$$
$$
=\text{sign}[\sin (\alpha \varphi /2)](r-R_m),
$$
\begin{equation}
\label{b14}
\qquad \varphi \in
[2\pi (n-m+1)/\alpha ,2\pi (n-m)/\alpha ],
\end{equation}
where $R_m$ is a position of the $m$th kink, $m=1,2,\ldots ,n$. When the
coordinate $r$ passes through the point $R_m$ the scalar field $\varphi (r)$
changes from $2\pi (n-m+1)/\alpha $ to $2\pi (n-m)/\alpha $ (see Fig. 1).
Eq. (\ref{e4}) yields the following expression for $g$ as a function of $%
\varphi $ inside the bubble:
\begin{equation}
\label{b15}\frac 1g=1-\frac{4R}{\alpha ^2}[2m-1+\cos (\alpha \varphi /2)].
\end{equation}
Outside the bubble $\varphi =0$, $m=n$ and $1/g=1-8nR/\alpha ^2$. The
solution is valid if $1/g>0$, that is $R<R_{\max }=\alpha ^2/8n$. Match of
the inner solution (\ref{e4}) with the Schwazschild solution (\ref{s1})
determines the mass-radius relation
\begin{equation}
\label{b16g}M=4\pi nuR^2-8\pi ^2n^2u^2R^3,
\end{equation}
where $u$ is the surface energy density given by an integral over one
potential period $u=\int \sqrt{2V}d\varphi /4\pi $. For the cosine potential
(\ref{a3}) $u=2/\pi \alpha ^2$.

\begin{figure}
\bigskip
\centerline{\epsfxsize=0.45\textwidth\epsfysize=0.33\textwidth
\epsfbox{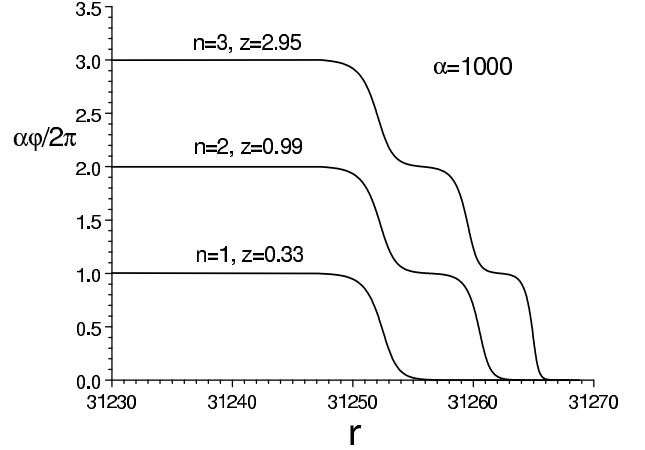}}

\caption{Scalar field $\varphi $ as a function of distance $r$ to the
bubble center for bubbles with equal radius and different quantum numbers $n=1$,
$2$, $3$. The unit of length is $\hbar/mc$. Note, we plot the field $\varphi $
only in the vicinity of the bubble surface where it undergoes variation.
}

\end{figure}

Redshift of the bubble interior $z=1/N-1$ can be found by matching the inner
$N=const$ and the outer (\ref{s1}) solutions:
\begin{equation}
\label{b17}z=\frac 1{\sqrt{1-2M/R}}-1=\frac 1{1-4\pi nuR}-1.
\end{equation}
The internal redshift monotonically increases from zero to infinity when the
bubble radius $R$ changes from zero to $R_{\max }$. Fig. 2 shows the
redshift of space as a function of the distance $r$ to the bubble center.
The redshift is constant in the bubble interior and monotonically decreases
outside the bubble.

\begin{figure}
\bigskip
\centerline{\epsfxsize=0.45\textwidth\epsfysize=0.33\textwidth
\epsfbox{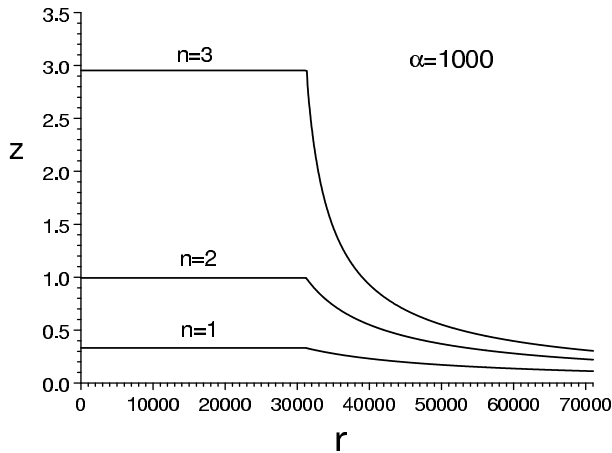}}

\caption{Redshift $z$ of space as a function
of distance $r$ to the bubble center for bubbles shown in Fig. 1.
}
\end{figure}

Let us make rescaling $M\rightarrow M/4\pi u$, $R\rightarrow R/4\pi u$, then
Eqs. (\ref{b16g}), (\ref{b17}) yield
\begin{equation}
\label{b18}M=nR^2-n^2R^3/2,\qquad z=\frac 1{1-nR}-1.
\end{equation}
For a given $M$ the redshift depends on the integer number $n$, which
implies the redshift is quantized.

In early samples of QSOs
associated with nearby spiral galaxies, Karlsson showed that the redshift
distribution has a periodicity $\log (1+z_{n+1})-\log (1+z_n)=0.089$, where $%
n=0,1,2,$ $\ldots $ and $z_0=0.061$ \cite{Karl71}. It has been later
confirmed by other groups \cite{Barn76,Arp90}. In a recent paper, Burbidge
and Napier \cite{Burb01} tested for the occurrence of this periodicity in
new QSO samples and found it to be present at a high confidence level. The
peaks were found at $z\approx 0.30$, $0.60$, $0.96$, $1.41$ and $1.96$ in
agreement with Karlsson's empirical formula. The formula also includes the
peak at $z_0=0.061$, however, this peak does not occur for quasars, but for
morphologically related objects.

The redshift periodicity is observed only in QSO samples satisfying certain
selection criteria, in particular, the galaxies which are assumed to be
paired to the QSOs must be {\it most nearby} spirals \cite{Karl90,Napi03}. This
implies that redshift quantization is a property of intrinsically faint QSOs
which are not detected from large distances.

It is naturally to assume that QSOs born in the same type of galaxies have
approximately equal masses because their formation mechanism must be
similar. Such phenomenon is well known for type Ia supernovae or
neutron stars: practically all
measured neutron star masses cluster around the value of $1.4M_{\odot }$
with only a few percent deviation \cite{Glen00}. If dark matter bubbles are
born with equal masses then, according to Eqs. (\ref{b18}), their redshift
must be quantized. For $M=0.0601$ (in dimension units $M=0.00752${\small $%
\alpha ^2m_{\text{pl}}^2/m$)} Eqs. (\ref{b18}) have solutions for $n=1,$ $2,$
$\ldots ,$ $8$, they are given in Table 1.

\begin{table}
\centerline{
\begin{tabular}{|c|c|c|}
\hline
n & R, in $\alpha^2\hbar/mc$ & z \\
\hline
1 & 0.0329 & 0.357  \\
2 & 0.0241 & 0.629  \\
3 & 0.0204 & 0.96  \\
4 & 0.0182 & 1.40  \\
5 & 0.0168 & 2.06  \\
6 & 0.0159 & 3.24  \\
7 & 0.0153 & 6.11  \\
8 & 0.0151 & 26.6  \\
\hline
\end{tabular}
}
\caption{Redshift of the bubble interior $z$ and
its radius $R$ for $M=0.00752\alpha ^2m_{\text{pl}}^2/m$ and different kink
numbers $n$. }
\end{table}

\begin{figure}
\bigskip
\centerline{\epsfxsize=0.45\textwidth\epsfysize=0.37\textwidth
\epsfbox{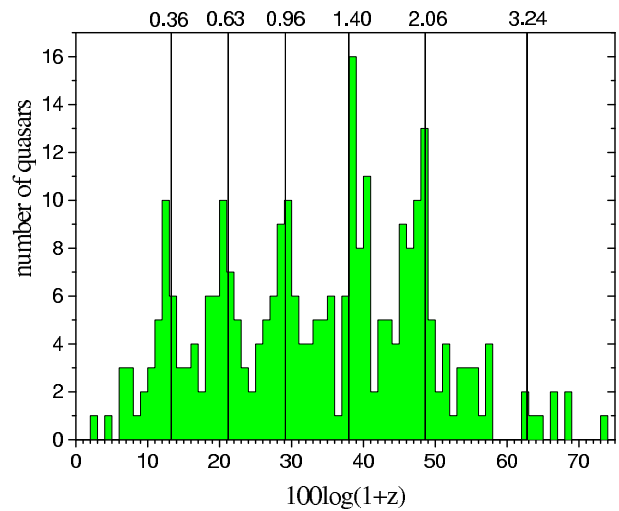}}

\caption{Histogram of the redshift distribution of QSOs close
to bright nearby
active spiral galaxies or multiple QSOs with small angular separation
from Ref. \cite{Napi03}.
The solid lines represent position of the peaks from Table 1.
}
\end{figure}

In Fig. 3 we plot the most recent histogram of the redshift distribution
from Ref. \cite{Napi03} in which five peaks are clearly seen. The solid
lines show the redshifts from our Table 1, they match well the observed
peaks. The agreement is remarkable because the theory has only one free
parameter, the bubble mass $M$. Such coincidence strongly suggests that the
point-like quasars associated with nearby galaxies are dark matter bubbles
composed of scalar particles with periodic interaction potential. One should
mention an alternative possibility of quasar evolution. Bubbles can be
originally born with the same mass and number of kinks $n=5$ that
corresponds to the $5$th peak. During evolution the kinks tunnel to the
bubble center and quasars sequentially decay into states with smaller $n$
but the same mass.

For axions with $m=0.1-3$ meV and
$f=2\times 10^9-6\times 10^{10}$ GeV Eq.
(\ref{a3}) yields $\alpha =5.6\times 10^7-1.7\times 10^9$. Hence, an axion
bubble with the internal redshift $z=0.36$ and $n=1$ would have the mass
$M=3\times 10^7-10^9M_{\odot }$
and the radius $R=3\times 10^2-10^4R_{\odot }$.
Such radius range agrees with the size of the
emission region expected for the intrinsically faint QSOs.
Indeed, for Seyfert 1 galaxies the size of the broad-line region is $R\sim
10-100$ light days. The luminosity $L$ of the QSOs is 5-6 orders
smaller. Based on the empirical relation for Seyfert 1 galaxies
$R\propto L^{1/2}$ \cite{Wang03}, we obtain for the quasars
$R\sim 10^4R_{\odot }$.

Now we discuss the bubble life time. Under the influence of surface tension
and gravitational attraction an initially static bubble starts to collapse.
In the thin-wall approximation the initial acceleration is given by \cite
{Blau87}
\begin{equation}
\label{bL1}\ddot R=-\frac{2N^3}R-\frac{N^2M}{(1+N)R^2},
\end{equation}
where $N=\sqrt{1-2M/R}$. For one kink thin-wall contracting bubble the
conserved mass is \cite{Blau87}
\begin{equation}
\label{bL2}M=\frac{4\pi uR^2}{\sqrt{1-(dR/d\tau )^2}}-8\pi ^2u^2R^3,
\end{equation}
where $\tau $ is the interior coordinate time. Based on Eqs. (\ref{bL1}), (%
\ref{bL2}) one can expect a continuous contraction of the bubble to the
origin on an astronomically short time scale $R/c\ll 1$yr. However, so far
we treated the scalar field as classical. Quantum corrections suppress the
collapse and result in appearance of long-lived bubbles, stable on a scale
more then
hundreds million years. To include quantum effects it has been suggested
that the expression (\ref{bL2}) be interpreted as the canonical hamiltonian
of the bubble at the quantum level \cite{Bere88,Auri90,Auri91}. The bubble
wave function $\Psi (R)$ satisfies the following stationary quantum
mechanical equation in one dimension ($\hbar =1$) \cite{Auri91}:
\begin{equation}
\label{bL4}\left[ \left( E+8\pi ^2u^2R^3\right) ^2+\frac{\partial ^2}{%
\partial R^2}-16\pi ^2u^2R^4\right] \Psi (R)=0.
\end{equation}
This equation possesses stationary solutions that are not possible in the
classical model.
Bubbles of non-negligible redshift correspond
to highly excited stationary states for which the energy spectrum can be
treated as quasi-continuous. At the quantum level the collapse is prevented
by quantum pressure that balances the surface tension and gravitational
attraction producing stationary configurations.

Let us estimate the decay time of an excited stationary state of the quantum
bubble. The decay occurs by means of scalar particle emission. We estimate
the decay time using the Bohr correspondence principle as the
time of energy loss by the classical bubble with the radius $R(t)$
oscillating between the turning points
$R(t)=R$ and $R(t)=0$, where $R$ is
determined by Eq. (\ref{b16g}). In the quantum picture, however, there are
no such oscillations. The probability of creation a particle with the energy
$mc^2$ by a moving bubble surface is governed by the Boltzmann factor $\exp
(-mc^2/T_{\text{eff }})$, where $T_{\text{eff }}=a/c$ is the effective
temperature and $a$ is the acceleration of the surface \cite{Gors00}. For
the bubble $a\approx c^2/R(t)$ and the Boltzmann factor reduces to $\exp
(-R(t)/l)$ where $l=\hbar /mc$ is the surface width. Hence, emission of
scalar particles is exponentially suppressed apart from small regions where $%
R(t)\lesssim l$. As a result, during one period of oscillation, $t_{\text{c}%
}\sim R/c$, the energy loss is $\Delta E\sim (l/R)E$, which yields the
bubble life time
\begin{equation}
\label{lt}
t\sim \frac{R}lt_{\text{c}}=\frac{R^2}{cl}.
\end{equation}
For an axionic bubble with $R>10^2R_{\odot }$ and $l<0.7$ cm Eq. (\ref{lt})
yields $t>10^8$ yrs which is the time we need to account for the
phenomenon of quasars.

Properties of the intrinsically faint point-like QSOs, combined with equations
for the bubble mass $M=0.00752\alpha ^2m_{\text{pl}}^2/m=2.94m
\text{(eV)}\times 10^{11}M_{\odot }$
and the radius $R=0.0329\alpha ^2\hbar/mc=
2.73m\text{(eV)}\times 10^{6}R_{\odot }$,
allow us to determine the axion mass $m$. The quasar luminosity
suggests that the bubble radius is larger then $10^3R_{\odot }$ which yields
$m>0.4$ meV and $M>10^8M_{\odot }$. From the other hand, the
quasar ejection from active galaxies implies that the bubble mass $M$ must
be much smaller then the galactic mass. It is reasonable to constrain $%
M<10^9M_{\odot }$ which leads to $m<3$ meV and $R<10^4R_{\odot }$%
. We conclude, the axion mass is $m=0.4-3$ meV. This value fits
in the open window for the axion mass constrained by astrophysical and
cosmological arguments \cite{Brad03}, which unambiguously points
towards the axionic nature of dark matter composing the intrinsically faint
point-like quasars. Current cavity search
experiments in Livermore \cite{Aszt02} and Kyoto University
\cite{Brad03} are looking for the axion in the mass range $1-10$ $\mu$eV which
deviates by two orders of magnitude from our result. Probably now, when the
axion mass is established from quasar observations, the axion has a better
chance to be discovered.

We mention that data on central ``black hole" masses in small companion galaxies
allow us to determine the axion mass more accurately and yield $m=1.0-1.9$ meV.
We will discuss this elsewhere. Moreover, observations show that apart from the
intrinsically faint point-like objects considered here there is a subgroup of
bright quasars which probably also possess noncosmological redshift.  Tachyons,
another dark matter candidate, can explain their nature.  We discuss this in a
detail paper \cite{Svid04}.

\end{document}